\documentclass[aps,twocolumn,superscriptaddress,footinbib,floatfix,showpacs]{revtex4} 
  
\usepackage{graphicx}

\begin{document} 
  
\title{Nonlinear and Nonequilibrium Dynamics in Geomaterials} 
      \vbox to 0pt{\vss 
                    \hbox to 0pt{\hskip-40pt\rm LA-UR-04-1122\hss} 
                   \vskip 25pt} 
\preprint{LA-UR-04-1122} 
\author{James A. TenCate} 
\affiliation{EES-11, University of California, Los Alamos 
National Laboratory, Los Alamos, New Mexico 87545} 
\author{Donatella Pasqualini} 
\affiliation{EES-11, University of California, Los Alamos 
National Laboratory, Los Alamos, New Mexico 87545} 
\author{Salman Habib} 
\affiliation{T-8, University of California, Los Alamos 
National Laboratory, Los Alamos, New Mexico 87545} 
\author{Katrin Heitmann} 
\affiliation{ISR-2, University of California, Los Alamos 
National Laboratory, Los Alamos, New Mexico 87545} 
\author{David Higdon} 
\affiliation{D-1, University of California, Los Alamos 
National Laboratory, Los Alamos, New Mexico 87545} 
\author{Paul A. Johnson} 
\affiliation{EES-11, University of California, Los Alamos 
National Laboratory, Los Alamos, New Mexico 87545} 
  
\begin{abstract}  
The transition from linear to nonlinear dynamical elasticity in rocks
is of considerable interest in seismic wave propagation as well as in
understanding the basic dynamical processes in consolidated granular
materials. We have carried out a careful experimental investigation of
this transition for Berea and Fontainebleau sandstones. Below a
well-characterized strain, the materials behave linearly,
transitioning beyond that point to a nonlinear behavior which can be
accurately captured by a simple macroscopic dynamical model. At even higher
strains, effects due to a driven nonequilibrium state, and relaxation
from it, complicate the characterization of the nonlinear behavior. 
\end{abstract} 
  
\pacs{62.40.+i, 62.65.+k, 91.60.Lj}  
  
\maketitle 
  
Rocks possess a variety of remarkable nonlinear elastic properties
including hysterisis with end-point memory~\cite{hyst}, variation of
attenuation and sound velocity with strain~\cite{winkler}, strong
dependence of elastic and loss constants on pressure, humidity, and
pore fluids~\cite{phum}, long-time relaxation phenomena (`slow
dynamics')~\cite{ts}, and nontrivial variation of resonance
frequency with strain~\cite{nonh,recdat}. Significantly, materials as
diverse as sintered ceramics and damaged steels are now known to
display similar effects~\cite{recdat}. Thus rocks may be viewed as
representative members of a class of fascinating, but poorly
understood, nonlinear elastic materials: Fundamental questions still
to be resolved relate not only to the underlying causes of the
nonlinear phenomena but also to the conditions under which they occur.
  
In this Letter, we focus on delineating two strain thresholds, one
below which the rocks behave effectively as linear elastic materials,
$\epsilon_L$, the other beyond which memory and conditioning effects
occur, $\epsilon_M$, and the dynamic elastic behavior straddling the
region of these thresholds. While in Ref.~\cite{winkler} it was argued
that $\epsilon_L \sim 10^{-6}$ (albeit with some uncertainty), more
recent data~\cite{recdat,nonh,gtj} have been used to support an
extension of the nonlinear region to substantially lower strains;
doubt has been cast even on the very existence of a
threshold~\cite{gj}. In addition, results from resonant bar
experiments~\cite{nonh,gtj} have been interpreted to exhibit a
`nonclassical' frequency and loss dependence on the drive amplitude,
i.e., frequency and $Q$ softening linearly with drive amplitude rather
than quadratically as predicted by Landau theory~\cite{ll}, even at
strains as small as $10^{-8}$. (The importance of $\epsilon_M$ in
interpreting resonant bar data is emphasized below.)
 
We have carried out a new set of well-characterized experiments, over
a wide dynamic range, to unambiguously settle these questions: While
longitudinal resonant excitation of bars is a classic measurement
technique~\cite{rbar}, rock samples require substantial care in terms
of controlling the temperature and humidity and characterizing
possible systematic effects, especially those due to conditioning of
the sample by the external drive~\cite{ts}.
  
Our major conclusions are as follows.  For Berea and Fontainebleau
sandstone samples, below a threshold strain $\epsilon_L\sim
10^{-8}-10^{-7}$ (lower end for Fontainebleau, upper end for Berea),
there was no discernible dependence of the resonance frequency on the
strain -- the materials behaved linearly to better than 1 part in
$10^4$. For $\epsilon > \epsilon_L$, the materials displayed a
reversible softening of the resonance frequency with strain, in
excellent qualitative agreement with the quadratic prediction of
classical nonlinear theory~\cite{ll} up to a point where memory and
conditioning effects became apparent ($\epsilon\sim\epsilon_M$). In
detailing and characterizing the onset and nature of the nonlinearity,
our results very substantially improve on previous
work~\cite{winkler}. We show below that, up to the conditioning
threshold, the dynamical behavior is accurately captured by a
phenomenological macroscopic model incorporating a (softening) Duffing
nonlinearity and linear losses. Beyond the conditioning threshold, the
simultaneous presence of nonlinearity and nonequilibrium dynamics
complicates the characterization of dynamical behavior; in the absence
of a separation of these effects, the data cannot be interpreted to
support the existence of nonclassical behavior.

\begin{figure}
\leavevmode\includegraphics[width=\hsize]{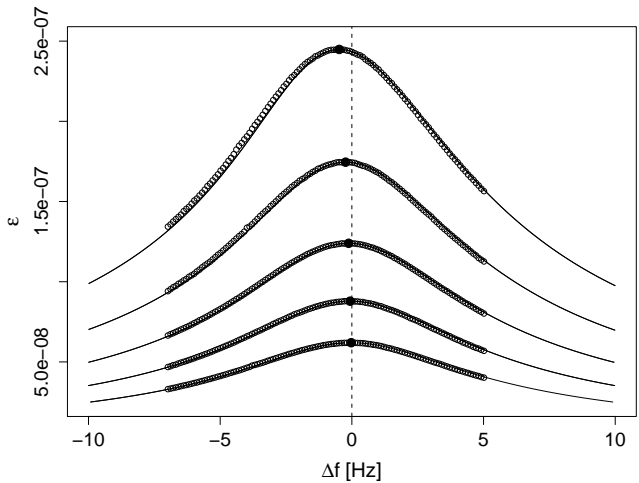}
\caption{The average strain amplitude $\epsilon$ as a function of
drive frequency for Fontainebleau sandstone. The reference center
frequency is 1155.98~Hz. The open circles are the experimental data;
the filled circles mark the peak positions. The solid lines are
theoretical predictions from Eqn.~(\ref{freq}).}  
\label{fig1}
\end{figure}

Our computer-controlled resonant bar experiment uses cylindrical
sandstone samples, 2.5~cm in diameter and 35~cm in length. The
cylinders are driven sinusoidally at one end by a PZT transducer with
a brass backload. The acceleration is measured by a B\&K accelerometer
at the other end of the bar and converted to an effective average
strain using the (known) driving frequency, $f$, via
$\epsilon=\ddot{u}/(4\pi Lf^2)$, where $u$ is the displacement and $L$
is the bar length. The finite accelerometer signal to noise restricts
the useful strain sensitivity to $\sim 10^{-10}-10^{-9}$ while the
upper end of strain is limited primarily by the physical integrity of
the experimental setup to $\sim 10^{-5}$. Scans of the resonance peak
are conducted at constant drive amplitude over an up/down frequency
sweep with spot measurements at frequencies $f_i$ bracketing the bar's
resonance frequency (Fig.~\ref{fig1}). While in previous
experiments~\cite{gtj} the temperature was actively controlled, the
present experiments were carried out using passive thermal isolation
to avoid even low-level thermal shocks.

The thermal history of a rock sample is known -- due to slow build-up
and relaxation of internal stresses -- to influence the effective
elastic modulus, and hence the resonance frequency of the bar. The
timescales associated with this behavior can be quite long, of the
order of several hours, hence long-term temperature stability is a
basic necessity for resonant bar experiments, especially at small
strain levels. With the present isolation system, long-term frequency
stability of the samples has been verified at $\sim 0.1$~Hz
(corresponding to a long-term thermal stability of $\sim 10$~mK),
which is close to how well the peak of the frequency response curve
can be determined at the lowest levels of strain reported in this
Letter.
 
The basic quantity measured in these experiments is the resonance
frequency, $f_0=\omega_0/2\pi$, of the bar as a function of the
strain, $\epsilon$, defined by the peak of the resonance curve as
measured above. The statistical analysis used a nonparametric Gaussian
process to model data trends. Bayesian estimation and characterization
of uncertainty was carried out using Markov chain Monte
Carlo~\cite{carlin}.

Traditionally, the loss, as represented by the $Q$ of the bar has also
been reported. However, this quantity is not easy to define or measure
precisely [the present bars have $Q\simeq 66$ (Berea) and $Q\simeq
132$ (Fontainebleau)] even at high values of $Q$~\cite{ibm}. We find
experimentally that the full-width of the response curves, $\Gamma$
(measured at $a_0/\sqrt{2}$, where $a_0$ is the peak amplitude;
$Q=\omega_0/\Gamma$) at all drive levels below the conditioning
threshold, is essentially invariant (this behavior is also predicted
by the macroscopic model described below). At higher drives, as
conditioning and memory effects appear, the response curves are not
symmetric around the peak and the width becomes strongly dependent on
the drive amplitude. This behavior will be detailed
elsewhere~\cite{us2}.
  
The determination of $f_0$ as a function of strain is complicated by
the fact that external driving can condition the sample and also lead
to shifts in the frequency via slow dynamics~\cite{ts}. In order to
eliminate this possibility we implemented a `zig-zag' strategy of
systematically increasing the drive level through the up/down
frequency sweeps, and then dropping back to the original drive
amplitude to verify that $f_0$ at the lowest strain value had not
changed. Application of this method shows that there are no
conditioning effects for $\epsilon < \epsilon_M\simeq 5\times10^{-7}$
for Berea and $\epsilon < \epsilon_M\simeq 2\times10^{-7}$ for
Fontainebleau. Up to these strain levels, any variation in $f_0$ as a
function of drive amplitude is taken to define the intrinsic
nonlinearity of the sample.
  
\begin{figure}
\leavevmode\includegraphics[width=\hsize]{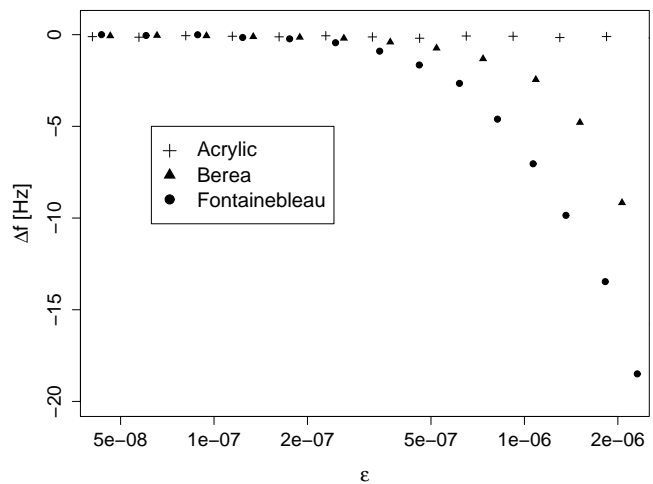}
\caption{Representative data for the resonant frequency shift $\Delta
f$ as a function of the effective strain $\epsilon$ for three
samples. Acrylic is a linear material used as a control in the
experiments. In this figure, the error bars are smaller than the sizes
of the symbols.}  
\label{fig2}
\end{figure}

\begin{figure}
\leavevmode\includegraphics[width=\hsize]{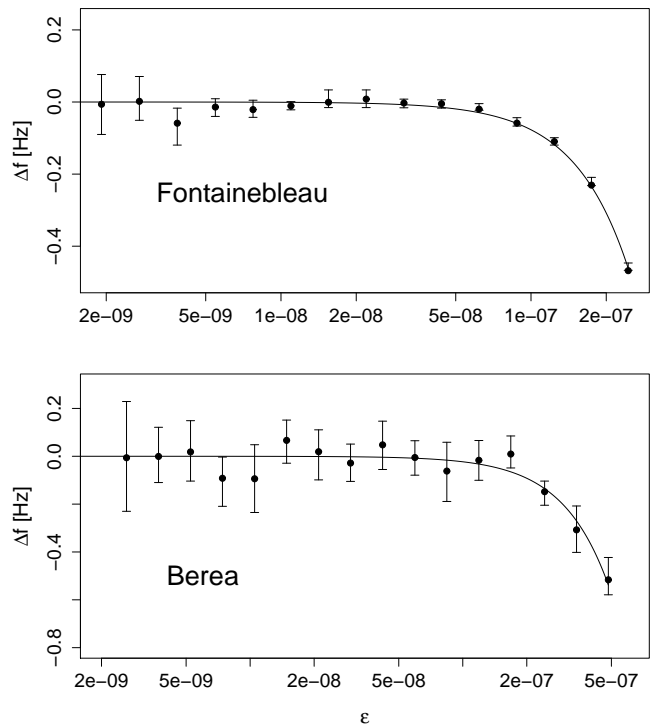}
\caption{The resonant frequency shift $\Delta f$ as a function of the
effective strain $\epsilon$ for Fontainebleau and Berea samples for
$\epsilon < \epsilon_M$. The solid lines represent predictions of the
theoretical model [Eqn.~(\ref{sig})]. Parameters for Fontainebleau are
$\Omega=7262.8$~rad/s, $\mu=27.5$~s$^{-1}$, $\gamma=-7.6\times
10^{19}$~m$^{-2}$s$^{-2}$, and for Berea, $\Omega=17375.7$~rad/s,
$\mu=131.6$~s$^{-1}$, $\gamma=-5.3\times 10^{19}$~m$^{-2}$s$^{-2}$.}  
\label{fig3}
\end{figure}

The measured variation of $f_0$ over a relatively wide range is shown
in Fig.~\ref{fig2} ($\Delta f$ is the difference between the measured
value and the starting, lowest drive level, value of $f_0$). Results
from a known linear material, Acrylic, are shown in order to provide a
control reference. Within the error bars of the experiment, the
materials are effectively elastically linear (no softening of $f_0$
with drive) till $\epsilon\sim 10^{-8}-10^{-7}$. Beyond this point,
$f_0$ softens quadratically with increasing drive amplitude until
$\epsilon\simeq\epsilon_M$ (Fig.~\ref{fig3}). We have compared our new
results with archival data from previous experiments on Berea reported
in Refs.~\cite{gtj}; allowing for differences in the starting
values of $f_0$, we find that the two datasets are in excellent
agreement.  This is quite remarkable, and encouraging from the point
of view of sample-independence, since the samples were subjected to
very different environmental conditions in the two cases. The
erroneous conclusion of nonclassical nonlinearity (linear dependence
of $f_0$ on $\epsilon$) reached in this earlier work was due to the
limited dynamic range in strain of the analyzed data ($<1$~decade)
versus the present measurements (3~decades): The present data are
consistent with a zero value for the cofficient of a linear fit to the
softening with strain; the corresponding values of Refs.~\cite{gtj}
are an order of magnitude too large to agree with the measurements.
  
We now describe a simple phenomenological dynamical model for driven
rods that provides an excellent description of the measured data. The
model is not directly extracted from a one-dimensional (nonlinear)
wave equation~\cite{knopoff}. Rather, the procedure follows
statistical mechanics-based modeling of degrees of freedom coupled to
dissipative channels~\cite{rwz,theory}. The model adds a quartic
(Duffing) softening nonlinearity to a harmonic potential (a more
complex model for the higher strain regime is described in
Ref.~\cite{theory}). The equation of motion for the displacement is
taken to be:
\begin{equation}  
\ddot{u}+2\mu\dot{u}+\Omega^2 u+\gamma u^3=F\sin(\omega t), 
\label{mod} 
\end{equation} 
where $\gamma < 0$. Since the displacement, $u$, is small, multiscale
perturbation theory can be used to solve Eqn.~(\ref{mod}) very
accurately~\cite{nayfeh}. For the case relevant to the experiment
($\omega\sim\Omega$), the solution is $u=a\cos(\omega
t+\phi)$. The phase $\phi$ is of no interest here, while the
relation between the amplitude of the oscillation, $a$, and the drive
amplitude, $F$, is given by  
\begin{equation}
\Omega^2\mu^2a^2+a^2\left(\sigma\Omega-{3\over
  8}a^2\gamma\right)^2={1\over 4}F^2  
\label{freq}
\end{equation} 
where $\omega\equiv\Omega+\sigma$. It is straightforward to show
that the peak of the response curve has the value,
$a_0=F/(2\mu\Omega)$, and occurs at the drive angular frequency
$\omega_0=\Omega+\sigma_0$, with
\begin{equation}
\sigma_0={3F^2\gamma\over 32\mu^2\Omega^3}.
\label{sig}
\end{equation}
Thus, the model predicts a quadratic softening of the frequency with
the drive amplitude $F$. In addition, by solving Eqn.~(\ref{freq}) for
$\sigma$, and then substituting $a=a_0/\sqrt{2}$, it is easy to show
that the width of any response curve is an invariant,
$\Gamma=2\mu$. 

With these results in hand, it is straightforward to determine model
parameters. As the model predicts, we have verified that $\Gamma$ as
measured from the experimental curves is constant within $1\%$ up to
the strain $\epsilon_M$, this immediately determining the damping
coefficient $\mu$. The (linear) resonant frequency $\Omega$ and the
nonlinearity parameter $\gamma$ now follow by fitting the experimental
data for $\Delta f$ as a function of the drive using Eqn.~(\ref{sig})
(Fig.~\ref{fig3}). As apparent from Fig.~\ref{fig1}, with these
parameters fixed as just described, the model predictions are in
excellent agreement with the experimental response curves.

It has been previously claimed that the absence of frequency softening
is not sufficient to rule out nonlinearity in rocks as harmonic
generation may exist even in the absence of a discernible frequency
shift~\cite{nonh}: At least in the materials studied here, this does
not occur. Our theoretical model assumes that the fundamental mode
dominates the response of the bar to external driving and no higher
harmonics are excited via mode-coupling. With the parameters fixed as
above, harmonic generation via the intrinsic nonlinearity of the model
is very weak, with all even harmonics suppressed, and with odd
harmonics typically $80$~dB below the fundamental -- lower than the
noise floor of the experiment, and, consistent with this prediction,
we did not observe any harmonic content in the signal even at the
highest drive levels.

\begin{figure}  
\leavevmode\includegraphics[width=\hsize]{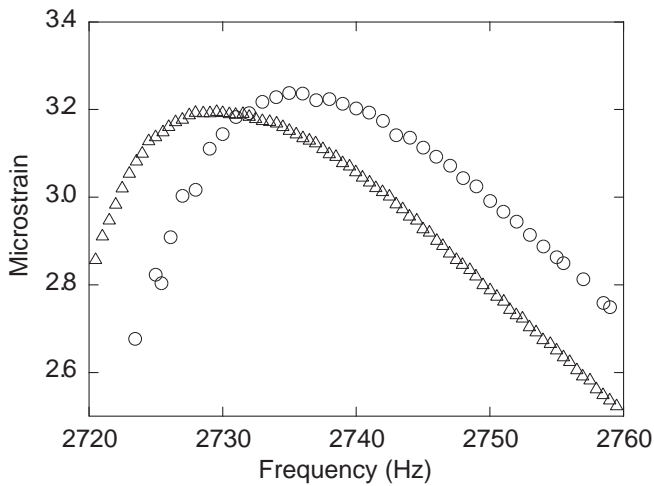} 
\caption{Avoiding relaxation effects: The left curve is for Berea
maintained in a nonequilibrium steady state by continuous driving,
while the right curve allows for a long relaxation to take place
between the individual data points.} 
\label{fig5} 
\end{figure} 
  
We now turn to the behavior of $f_0$ as $\epsilon$ is increased beyond
$\epsilon_M$, the point where conditioning begins to play a role. Here
the frequency softening is such that $f_0$ no longer returns to the
starting point when the drive is removed, but to a lower value. It is
clear that in this regime the measured softening at a given strain
cannot be interpreted as wholly due to an intrinsic nonlinearity; it
will as well depend on the sweep-rate~\cite{ts}. Analyzing a $f_0$
vs. $\epsilon$ curve without taking this effect into account would
introduce an unknown systematic effect exaggerating the actual
nonlinearity. Fortunately, experimental protocols can be implemented
to alleviate this difficulty as described in detail in
Ref.~\cite{us2}. To illustrate these issues, Fig.~\ref{fig5} shows
response curves for Berea sandstone in two different regimes. The left
curve represents an experiment where, at each frequency, the sample is
allowed to come to a new steady state (10 minutes at each
frequency), while the right curve allows for 10 minutes between
rapidly taken individual measurements to allow a return to thermal
equilibrium between measurements. In both cases, the measurement
protocols are designed to reduce relaxation effects as much as
possible, in one case by staying at all times in a nonequilibrium
steady state, and in the other, by allowing a close-to-equilibrium
return between measurements. The key point is that at strain levels
$\epsilon > \epsilon_M$, the rock transitions to a nonequilibrium
state characterized by a different set of macroscopic parameters, as
evidenced by the 5~Hz shift between the left and right curves. If
measurements could be made rapidly enough so that each point on the
right curve corresponds to an equilibrium state, then it is precisely
these response curves that are equivalent to data taken at $\epsilon <
\epsilon_M$. In practice, however, this procedure is very difficult to
carry out as it requires stringent long-term environmental
stability. Thus, all measurements to date in this higher strain regime
are dangerously contaminated by conditioning; the separate effects of
nonlinearity and relaxation in these experiments cannot be
disentangled.

To summarize, our experiments have established the existence of a
reversible dynamic quadratic nonlinearity in the Berea and
Fontainebleau sandstones up to a material-dependent strain threshold,
$\epsilon_M$. Below this threshold we find no evidence for
nonclassical behavior as reported
previously~\cite{nonh,gtj}. Frequency shifts in dynamical experiments
with $\epsilon > \epsilon_M$ do not have a simple interpretation due
to the existence of a driven nonequilibrium state with differing
macroscopic parameters; because of the competition of material
nonlinearity and conditioning and relaxation effects, present
experimental data cannot distinguish classical from nonclassical
effects in this regime. Experiments to do so are in progress.

We are indebted to Tim Darling, Robert Guyer, and Tom Shankland for 
stimulating discussions.


\begin{thebibliography}{99} \bibitem{hyst} N.G.W.~Cook and K.~Hodgson,
J.~Geophys.~Res. {\bf 70}, 2883 (1965); R.B.~Gordon and L.A.~Davis,
{\em ibid} {\bf 73}, 3917 (1967).
  
\bibitem{winkler} K.~Winkler, A.~Nur, and M.~Gladwin, Nature {\bf
277},  528 (1979) and references therein. 
  
\bibitem{phum} S.G.~O'Hara, Phys.~Rev.~A {\bf 32}, 472 (1985);
B.~Zinszner, P.A.~Johnson, and P.N.J.~Rasolofosaon,
J.~Geophys.~Res. {\bf 102}, 8105 (1997).
  
\bibitem{ts} J.A.~TenCate and T.J.~Shankland, Geophys. Res. Lett. {\bf
23}, 3019 (1996); J.A.~TenCate, E.~Smith, and R.A.~Guyer,
Phys. Rev. Lett. {\bf 85}, 1020 (2000).
  
\bibitem{nonh} P.A.~Johnson, B.~Zinszner, and P.N.J.~Rasolofosaon,
J.~Geophys.~Res. {\bf 101}, 11553 (1996). 
    
\bibitem{recdat} For a review, see L.A.~Ostrovsky and P.A.~Johnson,
Riv. Nuovo Cimento {\bf 24}, 1 (2001).

\bibitem{gtj} R.A.~Guyer, J.~TenCate, and P.A.~Johnson,
Phys. Rev. Lett. {\bf 82}, 3280 (1999); E.~Smith and J.A.~TenCate,
Geophys. Res. Lett. {\bf 27}, 1985 (2000).
  
\bibitem{gj} R.A.~Guyer and P.A.~Johnson, Physics Today {\bf 52}, 30
(1999).  
  
\bibitem{ll} L.D.~Landau and E.M.~Lifshitz, {\em Theory of 
Elasticity}, (Butterworth-Heinemann, Boston, 1998). 
  
\bibitem{rbar} S.P.~Clark (Ed.) {\em Handbook of Physical Constants}
(Geological Society of America, New York, 1966), J.C.~Jaeger and
N.G.W.~Cook, {\em Fundamentals of Rock Mechanics} (Chapman and Hall,
London, 1979); R.S.~Carmichael (Ed.), {\em CRC Handbook of Physical
Properties of Rocks} (CRC Press, Boca Raton, 1984); T.~Bourbie,
O.~Coussy, and B.~Zinszner, {\em Acoustics of Porous Media} (Gulf,
Houston, 1987).
  
\bibitem{carlin} S.~Banerjee, B.P.~Carlin, and A.E.~Gelfand, {\em
Hierarchical Modeling and Analysis for Spatial Data} (Chapman and
Hall/CRC Press, Boca Raton, 2004).
    
\bibitem{ibm} See, e.g., B.C.~Stipe, H.J.~Mamin, T.D.~Stowe,
T.W.~Kenny, and D.~Rugar, Phys. Rev. Lett. {\bf 87}, 096801 (2001). 

\bibitem{us2} J.~TenCate et al, in preparation.  
  
\bibitem{knopoff} See, e.g., L.~Knopoff and G.J.F.~MacDonald,
Rev. Mod. Phys. {\bf 30}, 1178 (1958). 

\bibitem{rwz} R.W.~Zwanzig, J. Stat. Phys. {\bf 9}, 215 (1973). 

\bibitem{theory} S.~Habib, K.~Heitmannn, and D.~Pasqualini, in
preparation.
  
\bibitem{nayfeh} A.H.~Nayfeh, {\em Introduction to Perturbation
Techniques} (Wiley, New York, 1981); G.~Schmidt and A.~Tondl, {\em
Non-Linear Vibrations} (Cambridge University Press, New York, 1986).

\end{thebibliography}
\end{document}